\documentclass{book}
\usepackage{amsmath}
\usepackage{amstext}
\usepackage{amssymb}
\usepackage{science}
\usepackage{graphicx}  

\newcommand{\beq}{\begin{equation}}
\newcommand{\eeq}{\end{equation}}
\newcommand{\bea}{\begin{eqnarray}}
\newcommand{\eea}{\end{eqnarray}}

\def\laq{~\raise 0.4ex\hbox{$<$}\kern 
-0.8em\lower 0.62ex\hbox{$\sim$}~}
\def\gaq{~\raise 0.4ex\hbox{$>$}\kern 
-0.7em\lower 0.62ex\hbox{$\sim$}~}

\def \ra {\rightarrow}
\def \noi {\noindent}
\def \la {\lambda}

\def \Da {\Delta}

\def \Sg {\Sigma}

\def \r {\rho}

\def \ls {\lambda_{\rm s}}
\def \lp {\lambda_{\rm P}}

\def \Mp {M_{\rm P}}

\def \d {{\rm d}}
\def \e {{\rm e}}

\begin{document}

\chapter*{String theory and primordial cosmology}

\begin{flushright}
BA-TH/685-14\\
\end{flushright}

\noi
M. GASPERINI\\

\noi
{\em Dipartimento di Fisica, Universit\`a di Bari, 
and Istituto Nazionale di Fisica Nucleare, Sezione di Bari, 
Via G. Amendola 173, 70126 Bari, Italy.} \\
E-mail: {\tt{gasperini@ba.infn.it}}\\

\index{string theory}
\index{string cosmology}

\noi
SUMMARY. \\ 

\noi
String cosmology aims at providing a reliable
description of the very early Universe in the regime where standard-model physics is no longer appropriate, and where we can safely apply the basic  ingredients of superstring models such as dilatonic and axionic forces, duality symmetries, winding modes, limiting sizes and curvatures, higher-dimensional interactions among elementary extended object. The sought target is that of resolving (or at least alleviating) the big problems of standard and inflationary cosmology like the spacetime singularity, the physics of the trans-Planckian regime, the initial condition for inflation, and so on. 

\begin{center}
\vspace{1.5cm}

------------------------------------------

\vspace{0.5cm}

{\large Published in {\em ``Springer Handbook of Spacetime''}\\

(ed. by A.  Ashtekar and V. Petkov),  Chapter 35

\bigskip

Springer-Verlag, Berlin Heidelberg, 2014}

\bigskip
ISBN 978-3-642-41991-1
\end{center}

\chapter{String theory and primordial cosmology}

M. GASPERINI\\
\noi
{\em Dipartimento di Fisica, Universit\`a di Bari, 
and Istituto Nazionale di Fisica Nucleare, Sezione di Bari, 
Via G. Amendola 173, 70126 Bari, Italy.} \\
E-mail: {\tt{gasperini@ba.infn.it}}\\

\index{string theory}
\index{string cosmology}

\noi
SUMMARY. String cosmology aims at providing a reliable
description of the very early Universe in the regime where standard-model physics is no longer appropriate, and where we can safely apply the basic  ingredients of superstring models such as dilatonic and axionic forces, duality symmetries, winding modes, limiting sizes and curvatures, higher-dimensional interactions among elementary extended object. The sought target is that of resolving (or at least alleviating) the big problems of standard and inflationary cosmology like the spacetime singularity, the physics of the trans-Planckian regime, the initial condition for inflation, and so on. 

\section{The standard ``Big Bang" cosmology}

In the second half of the last century the theoretical and observational study of our Universe, grounded on one hand on the Einstein theory of  general relativity, and on the other hand on astronomical observations of every increasing precision, has led to the formulation (and to the subsequent completion) of the so-called {\em standard cosmological model} (see e.g. \cite{1,2,3}). 
\index{standard cosmology}

Such a model -- like every physical model -- is based on various assumptions. We should mention, in particular:

Using such assumptions, the standard cosmological model has obtained a long and impressive series of successes and experimental confirmations, such as:

In spite of these important achievements the standard cosmological model was put in trouble when, in the $1980$s, the scientific community started to investigate the problem of the origin of the observed galactic structures, and of the small (but finite) inhomogeneity fluctuations presents in the temperature $T$ of the relic background radiation ($ \Da T/T \sim 10^{-5}$). How did originate the temperature inhomogeneities $ \Da T/T$ and, especially, the matter inhomogeneities $ \Da \r/\r$ which are at the grounds of the concentration and subsequent growth of the cosmic aggregates (cluster of galaxies, stars and planets) that we presently observe? No temperature fluctuation and density fluctuation should exist, on macroscopic scales, if our Universe would be exactly homogeneous and isotropic as required by the standard cosmological model. 
\index{temperature anisotropy}

This problem was solved by assuming that the standard cosmological model has to  be modified, at some very early epoch, by the introduction of a cosmological phase -- called {\em inflation} -- characterized by an accelerated expansion rate \cite{4,5,6}.  During such a primordial inflationary phase the three-dimensional spatial sections of our Universe underwent a gigantic (almost exponential) growth of proper volume in few units of the Hubbe-time parameter (see see.g. \cite{3,7,8}). \index{inflation}
This process was able to amplify the microscopic quantum fluctuations of the matter fields (and of the geometry), thus producing the macroscopic inhomogeneities required for the formation of the matter structures and of the temperature anisotropies we observe today (see see.g. \cite{8,9,10}).

\index{eternal inflation}
A phase of inflationary evolution like that proposed above, however, cannot be extended back in time to infinity  (or, to use the standard terminology, cannot be {\em past eternal} \cite{11,12,13}). If we go  back in time to sufficiently earlier epochs we find that the inflationary phase of the standard model has a beginning at a precise instant of time. Before that time, the Universe was in an extremely hot, dense and curved primordial state -- an ultimate concentrate of matter and radiation at extremely high energy and temperature.

\index{big bang epoch}
This means, in other words, that before starting inflating the Universe was quite close to the so-called {\em big bang} epoch, namely to the epoch of the huge cosmic explosion which -- according to the standard model, even including the inflationary phase -- gave rise to the matter and energy species we observe today, and was at the origin of the spacetime itself. 

In fact, the big bang epoch of the standard model corresponds (strictly speaking) to a mathematical singularity where the energy density and the spacetime curvature blow up to infinity. We can thus say that to the question {\em How did the Universe begin?}, the standard cosmological model provides the answer: {\em The Universe was born from the initial big bang singularity}.
\index{big bang singularity}

\subsection{Validity restrictions of the standard cosmological model}

It is well known that standard cosmology is based on the Einstein theory of general relativity, which is a {\em relativistic} theory of gravity, but {\em not} a {\em quantum} theory. Hence, like all classical theories, general relativity has a limited validity range. Because of those limits the standard cosmological model cannot be extrapolated to physical regimes where the energy and the spacetime curvature are too high: this prevents taking too seriously the predictions of such a model about the initial singularity. 

We should recall, in fact, that a classical model is valid until the corresponding action $S=Et$ is much larger than the elementary {\em quantum of action} (or Planck's quantum) $h$. If we take a cosmological patch of the size given by the Hubble radius $c/H$, we can then estimate the total involved energy $E$ by multiplying the energy density $\r$ of the gravitational sources by the spatial volume $(c/H)^3$, containing the contribution of all observable matter and radiation at a given time $t$. \index{Hubble radius} 
The typical cosmological time scale, on the other hand, is provided by the Hubble time $H^{-1}$, and the energy density $\r$ is related to the Hubble time by the Einstein equations, which imply (modulo numerical factors of order one) $\r = c^2 H^2/G$, where $G$ is the Newton gravitational constant. \index{Hubble time} 
By imposing the condition $Et \gg h$  we then find that the standard cosmological model may give a reliable (classical) description of the Universe provided that
\beq
{c^5\over G H^2} \gg h.
\label{1}
\eeq
(This condition, in units $h=c=1$, can also be rephrased as $H \ll \Mp$, where
$\Mp=(hc/G)^{1/2}$ is the Planck mass).
\index{Planck mass}

\begin{figure}[t]
\centering
\includegraphics[height=7cm]{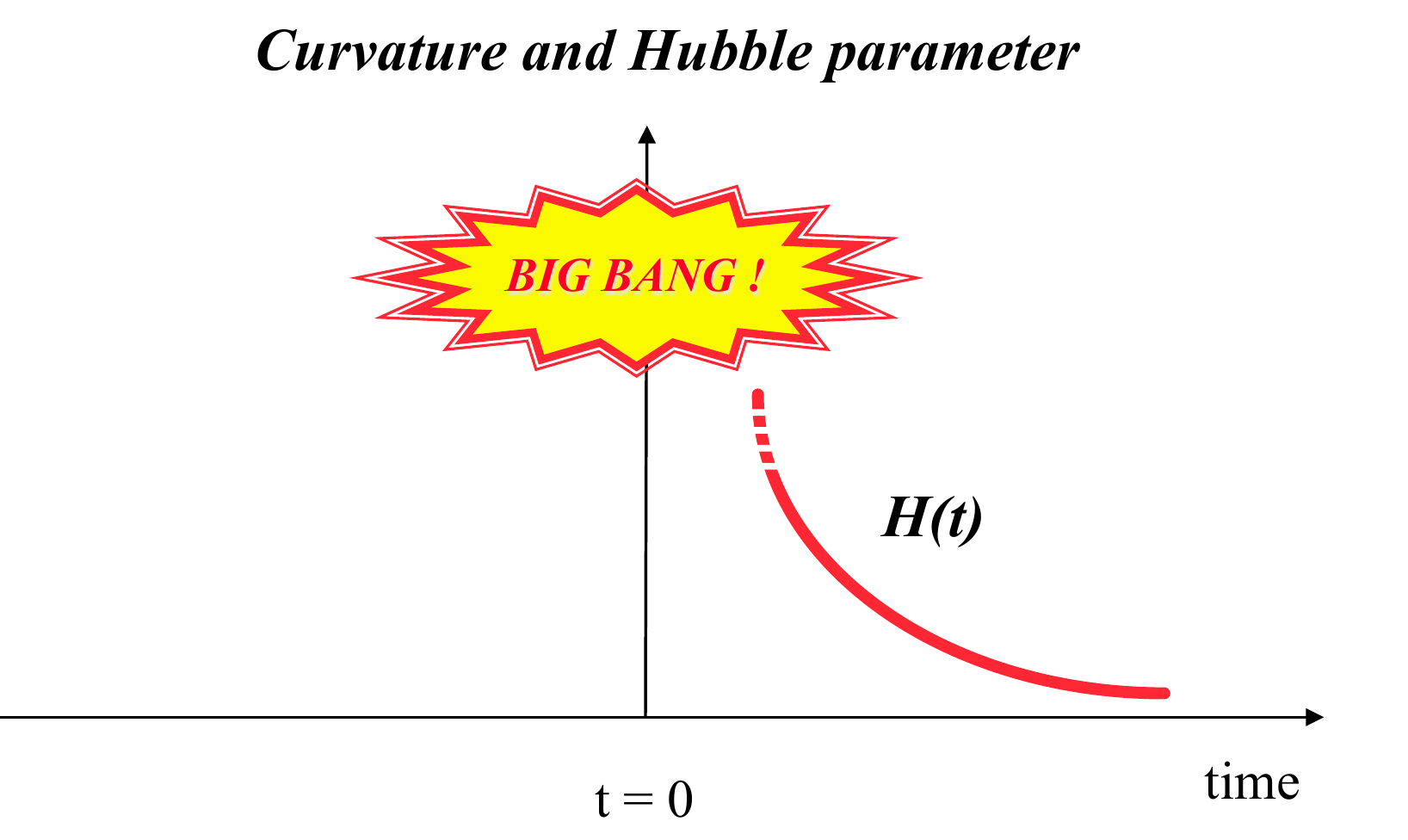}
\caption{\em According to the standard cosmological model, the spacetime curvature and the associated Hubble parameter $H(t)$ undergo an unbounded growth as we go back in time, and blow up at the time $t=0$ of the initial ``big bang" singularity.}
 \label{fig1}
\end{figure}

The parameters $C$, $G$, $h$ appearing in the above equation are constant, while the Hubble parameter $H$ is closely related to the spacetime curvature and is time dependent, $H=H(t)$. According to the standard model, in particular, $H$ grows as we go back in time, and diverges at the time of the big-bang singularity (Fig. \ref{fig1}). Correspondingly, the ratio $c^5/G H^2$ decreases and goes to zero at the singularity. Hence, before reaching the big bang epoch we necessarily enter the regime where the condition (\ref{1}) is violated, and the standard cosmological model is no longer valid.
\index{big bang singularity}

In order to provide a reliable description of the primordial Universe we should thus use a more general approach, based on a theory able to describe gravity also in the quantum regime. A possible candidate for this theory, which is complete, consistent at all energy scales and, besides gravity, also incorporates all fundamental interactions, is the so-called {\em theory of strings} (see, e.g., \cite{14,15,16}). 

\section{String theory}

\index{string theory}
The name of this theory is due to the fact that it proposes a model where the fundamental {\em building blocks} of our physical description of nature are one-dimensional extended object (elementary ``strings", indeed), instead of elementary particles. Such strings can be open (of finite length), or closed, and the spectrum of states associated to their vibration modes can reproduce the particle states of the gravitational interaction and of all the other fundamental (electromagnetic, strong and weak) interactions.
\index{open string}
\index{closed string}

In addition, if the string model is appropriately {\em supersymmetrized} -- namely, if we add to each bosonic degree of freedom a corresponding fermionic partner -- we arrive at the so-called theory of {\em superstrings}. This model potentially describes not only all interaction fields, but also their elementary sources (quarks and leptons), and thus all possible species and states of matter \cite{14,15}.
\index{superstring theory}

But there is more. A basic property of string theory -- probably the most revolutionary property, comparing with the other theories -- is the property of determining not only the possible form of the interaction terms (which is also done by the usual gauge theories, through the minimal coupling procedure), but also the form of the free-field (kinetic) terms (which in the other theories is always left, to some extent, arbitrary). Indeed, string theory satisfies a new symmetry (called {\em conformal symmetry}) which rigidly prescribes the allowed free-field dynamics, at any given order of the chosen perturbative expansion \cite{14,15}.
\index{conformal symmetry}

Quantizing a string, and imposing that the conformal symmetry is left unbroken by the quantum corrections (i.e. imposing the absence of {\em conformal anomalies}), one finds, in fact, that -- to lowest order -- the electromagnetic field {\em must} satisfy the Maxwell equations, the gravitational field {\em must} satisfy the Einstein equations, the spinor fields {\em must} satisfy the Dirac equations, and so on. All field equations, laboriously discovered in the past centuries through the theoretical elaboration of a large amount of empirical  data, can  be simply {\em predicted} by string theory even in the absence of any experimental input!

Finally, as already stressed, string theory is valid for all interactions also in the quantum regime, and can thus be used at arbitrarily high energy scales. In particular, unlike general relativity, can be applied to describe the Universe at epochs arbitrarily near to the big bang epoch. In such a limiting high-energy regime the equations we obtain from string theory are different, in general, from the corresponding field-theory equations, and thus it makes sense to ask the question {\em What's new from string theory about cosmology?}

In particular,  {\em What's new about the very early epochs at the beginning of the Universe?}

\section{String cosmology}

There are, in particular, two aspects of string theory which can play a relevant role in the formulation of a consistent cosmological scenario.
\index{string cosmology}
\index{duality simmetry}

\begin{figure}[t]
\centering
\includegraphics[height=7cm]{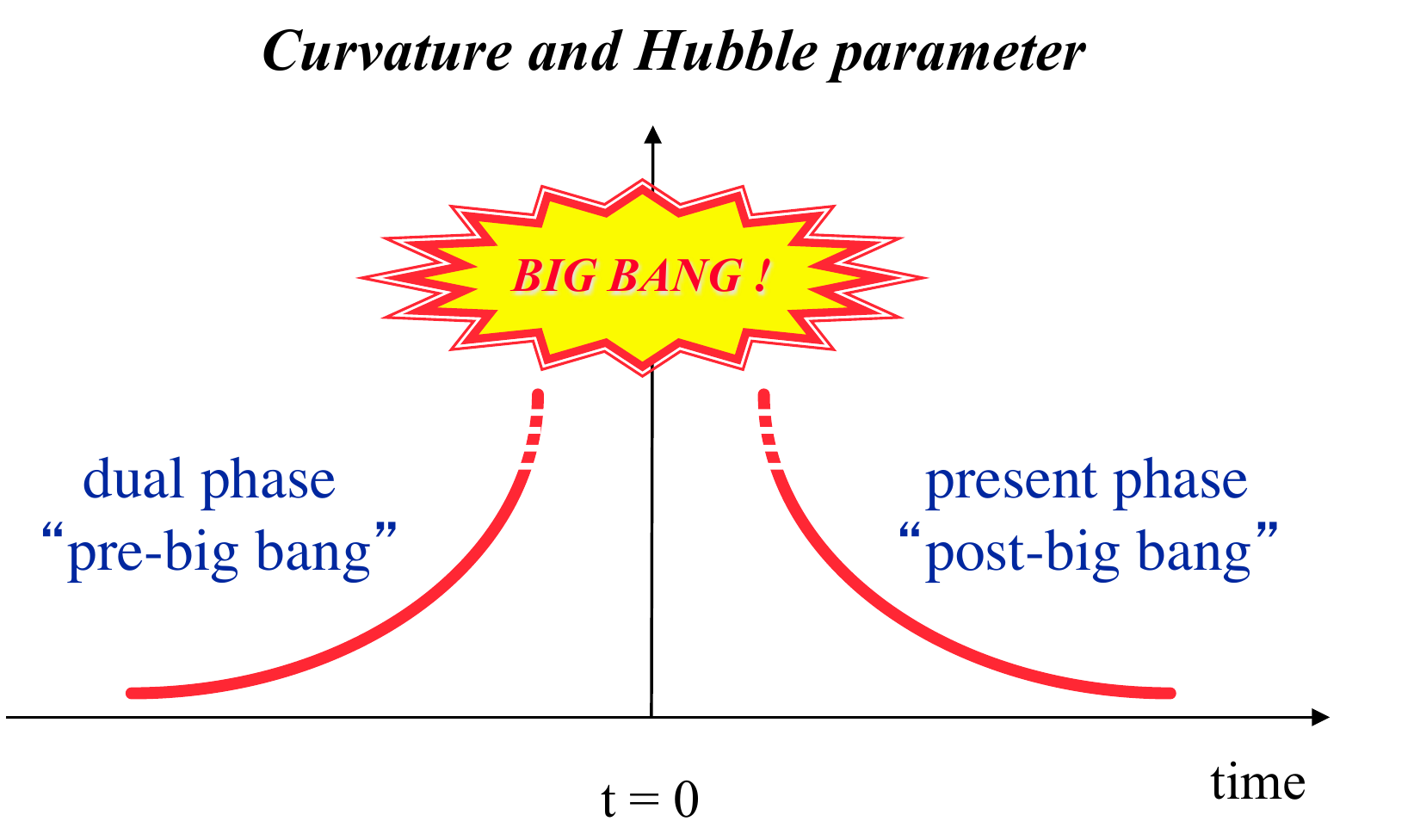}
\caption{\em The standard cosmological phase, of ``post-big bang" type, is preceded by a (string-theory) dual phase of ``pre-big bang" type.}
 \label{fig2}
\end{figure}

The first one concerns the so-called {\em dual symmetry}, typical of one-dimensional extended objects. If such a symmetry is respected (even at the approximate level) by the gravitational dynamics on cosmological scales, then any cosmological phase occurring at $t>0$, and characterized by a decreasing Hubble parameter $H$ (hence, decreasing curvature), must be associated to a {\em dual} partner phase, defined at $t<0$ and characterized by  growing $H$ (see \cite{17} for a nontechnical illustration of this duality symmetry). It follows in particular that the present cosmological phase, subsequent to the big bang epoch and well described by the standard model, must be preceded in time by an almost specularly symmetric phase occurring {\em before the big bang} (Fig. \ref{fig2}). Such a duality symmetry should also leave an imprint on the properties of the cosmological perturbations \cite{17a}.

\index{pre-big bang}
In Fig. \ref{fig2} both phases are characterized by a curvature (and a Hubble parameter) which diverges as $t$ goes to zero. If that would be the case, then the two branches of the cosmological evolution would be causally disconnected by a spacetime singularity, with no chances of merging together into a single coherent model of spacetime evolution. It is here, however, that comes into play another crucial aspect of string theory. 

\index{string length}
String theory is indeed characterized by a fundamental length $\ls$, which is a constant parameter of the string action and which controls the typical size of a quantized string. The physical role played by $\ls$ is very similar to the role played by the Bohr radius for the atom, which represents the minimal allowed size of the quantum electronic orbitals. The numerical value, however, is quite different: we may expect, in fact, $\ls \sim 10^{-33}$ cm (i.e.  a value of $\ls$ which is about $10$ times that of the Planck length $\lp= h/\Mp c^2$), in order that string theory may include a realistic description of all fundamental interactions (different values of $\la_s$ are possible in the presence of large extra dimensions, see below).
\index{Planck length}

Aside from the particular numerical value of $\ls$, what is important, in our context, is that proper distances and sizes smaller than $\ls$ have no physical meaning in a string-inspired model. It follows that, in a string-cosmology context, the Hubble radius $c/H$ has to be constrained by the condition $c/H \gaq \ls$. Since the Hubble radius is directly related to the inverse of the spacetime curvature, we can deduce that the curvature cannot blow up to infinity, because of the constraint $H \laq c/\ls$. Hence, when a given spacetime region has reached the limiting value $H \approx c/\ls$, its geometrical state can only evolve in two ways: it can either stabilize  at such a maximum value, or start decaying towards lower curvature states after a bounce induced by appropriate ``stringy'' effects (see e.g. \cite{17b}).  
\index{maximal curvature}
\index{string phase}

\begin{figure}[t]
\centering
\includegraphics[height=5cm]{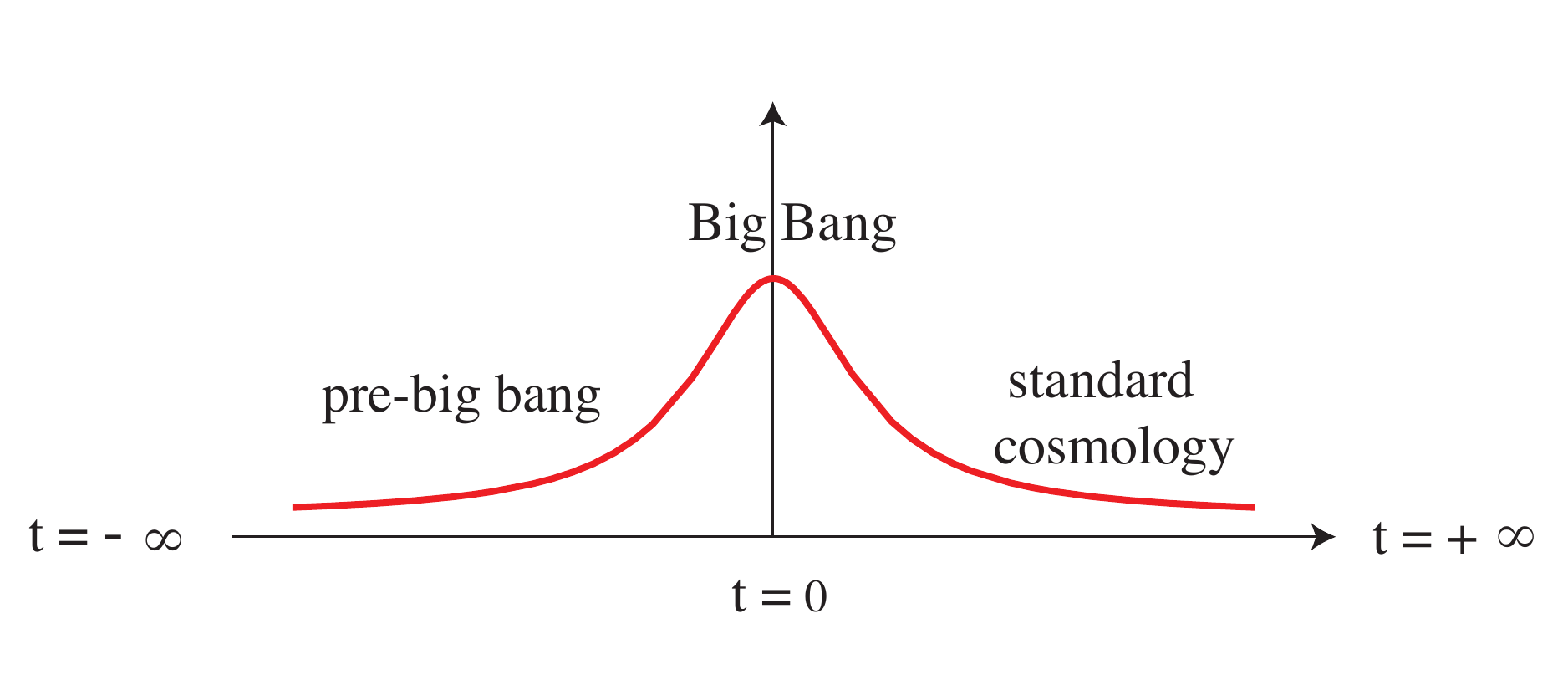}
\caption{\em Time evolution of the curvature scale and of the energy density in a typical example of string-cosmology scenario. The big bang epoch does not correspond to a singularity (like in the standard model) but to a phase of maximal, finite curvature. The Universe evolves starting from a flat, cold and empty state called the ``string perturbative vacuum", asymptotically localized at $t=-\infty$.}
 \label{fig3}
\end{figure}

In such a context, the big bang singularity predicted by the standard model and sharply localized at a a given epoch (say, $t=0$), has thus to be replaced by an extended phase of very high (but finite) maximal curvature: the so-called {\em string phase} (see e.g. \cite{18,19}). 
By combining the existence of the dual symmetry and of a minimal length scale, a string-based model can thus complete the standard cosmological scenario by removing the curvature singularity and extending the physical description of the Universe back in time, beyond the big bang, to infinity. The {\em big bang era} is still there, but it is deprived of the standard role of initial singularity: it corresponds, instead, to the  epoch marking the transition between the growing curvature and the decreasing curvature regime (Fig. \ref{fig3}). 

\index{string perturbative vacuum}
Within such a cosmological scenario (first presented in detail in \cite{20}) the initial cosmological state is no longer localized at $t=0$, but it is moved to the limit $t \ra - \infty$, and corresponds to an asymptotic state usually called the {\em string perturbative vacuum}. Such a new initial state, as  illustrated in Fig. \ref{fig3}, turns out to be a sort of specularly symmetric version of the final state that would be reached in the asymptotic future by a Universe which keeps expanding for ever according to the standard cosmological dynamics. Namely,  a flat, empty and cold initial state, drastically different from the initial hot, explosive state, extremely curved and concentrated, proposed by the standard scenario. 

There is, however, a  possible asymmetry between the initial and final state of the above  string-cosmology model, due to the coupling strength of the fundamental interactions: such a coupling tends to zero as $t \ra -\infty$, while it may become very strong in the opposite limit, if not appropriately stabilized (see e.g. \cite{18,19}). This growth of the coupling can be accompanied, in principle, also by a large amount of entropy production (see e.g. \cite{20a}).

\index{pre-big bang}

\section{A higher-dimensional Universe}

String theory, which is at the grounds of the cosmological scenario described in the previous section, can be consistently formulated only in the context of a higher dimensional spacetime manifold.

\index{ghost states}
\index{tachyon states}
In fact, in order to consistently quantize a bosonic string without introducing {\em ghosts} (states of negative norm), and without violating the Lorentz symmetry, one must introduce a generalized spacetime manifold with $26$ dimensions (see e.g. \cite{14,15}). In this way, however, one obtains a model which has still a pathology, as it contains {\em tachyons} (states of imaginary mass), which we believe should be absent in any realistic physical model. 

\index{superstring theory}
In order to eliminates the tachyons, we can generalize the bosonic string model by adding fermion states and considering the so-called {\em superymmetric} string models, or {\em superstring} models. In that case, a consistent quantization requires $10$ spacetime dimensions, and the number of total dimensions is to be increased up to $11$ (with one time-like and $10$ space-like dimensions) if we require that the five possible types of superstrings may be connected by duality transformations, and may represent various weak-coupling regimes of a more fundamental theory, called {\em M-theory} \cite{21} (see also \cite{22}). 
\index{M-theory}

Hence, whatever string model is assumed to apply, it is clear that the associated string cosmology scenario must be referred to a higher dimensional Universe. On the other hand,  all present phenomenological experience (including the most sensitive high-energy experiments) points at a world with {one} time-like and  {\em only three} space-like dimensions. We are thus naturally led to the following questions: 

{\em If string theory is correct, and the Universe in which we live has a number $d>3$ of spatial dimensions, why our experience is only limited to a three-dimensional space? why we cannot detect the additional ``extra" dimensions? what happened to those dimensions, if they really exist?}
\index{extra dimensions}

There are at least two possible answers to the above questions.

\index{Kaluza-Klein model}
There is an {\em old-fashioned} answer -- which, for a long time, has been also the only possible answer to the previous questions -- dating back to the so-called Kaluza-Klein model, formulated at the beginning of the last century \cite{23,24} in the context of a higher dimensional version of general relativity. According to such model, we cannot detect the extra dimensions simply because such dimensions  are compactified on length scales of  extremely small size (hence, they need extremely high energies in order to be experimentally resolved).

We can take, as a simple example, a long and very thin cylinder. A cylinder is a two-dimensional object but, if it is observed from a distance much larger than its radius, it may appear (in all respects) as being one-dimensional, extended in length but deprived of any sensible thickness. In the same way the spatial extension of our Universe could be largely asymmetric, with three spatial dimensions macroscopically expanded on a large scale, while all the other dimensions  {\em rolled up} in a highly compact way, and confined on a very small length scale -- of order (for instance) of $\ls$. If we do not have a sufficiently powerful instrument, able to resolve the required (very tiny) distance scales, we will always observe  three spatial dimensions only. 

\index{branes}
Very recently, however, a new possible answer to the dimensionality problem has been suggested by theoretical studies mainly performed in the second half of the $1990$s, and closely related to particular string-model configurations, called {\em branes} \cite{22}. Such a new answer states that we cannot ``see" the extra dimensions simply because the fundamental interactions propagate only along three spatial dimensions. All instruments we use to explore the world around us (starting from our eyes up to the more powerful and sophisticated technological tools) have indeed a working mechanism based on the fundamental (electromagnetic, nuclear, and so on) interactions. If such interactions are living only on a restricted subspace of the full spacetime manifold  (like, for instance, waves which propagate on the surface of a pond, and not in the direction orthogonal to the pond surface), then the extra dimensions are hidden to our direct experience, even if they are largely (or infinitely) extended. 

This second possible answer to the dimensionality problem has suggested new, interesting types of cosmological models, formulated in the context of the so-called {\em brane-world} scenario  (see, e.g., \cite{25}).

\section{Brane cosmology}

\index{brane-world cosmology}
According to the so-called brane-world cosmology, our Universe could be  a four-dimensional ``slice" of a higher-dimensional {\em bulk} manifold. The elementary charges sourcing the gauge interactions are confined on a three-dimensional hypersurface $\Sg_3$ associated to an object called  {\em Dirichet $3$-brane} (or $D_3$-brane), and we cannot detect the external spatial dimensions because the  gauge fields of those charges can propagate only on the {\em world-volume} $\Sg_4$ swept by the time evolution of the brane. (It should be recalled that the description of our Universe as a four-dimensional ``domain wall" embedded in a higher-dimensional bulk spacetime was previously suggested, with different motivations, also in \cite{26}). 
\index{Dirichlet brane}

\begin{figure}[t]
\centering
\includegraphics[height=7cm]{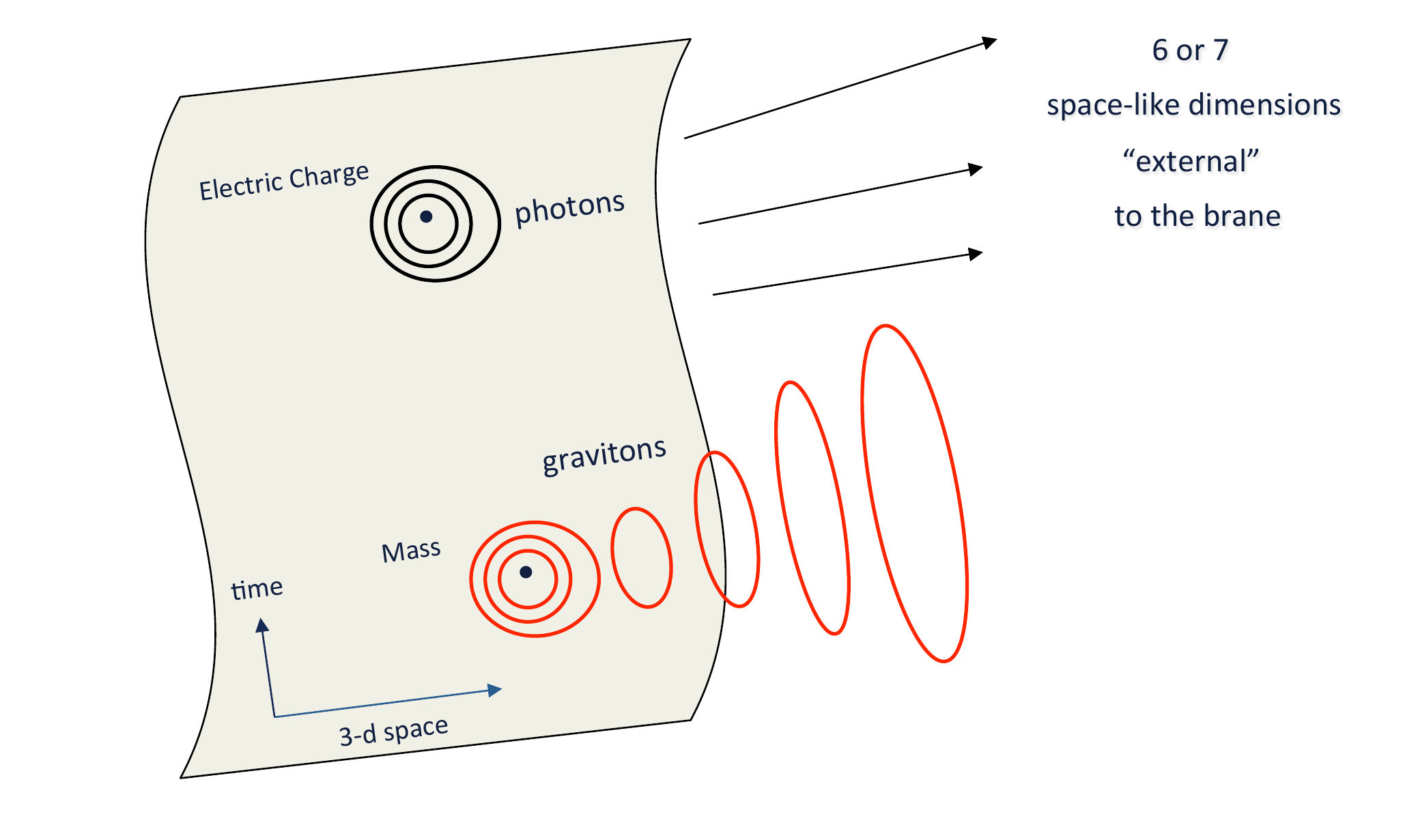}
\caption{\em A brane-Universe with one time-like and three space-like dimensions, embedded in an external bulk spacetime characterized by six (according to superstring theory) or seven (according to M-theory) extra spatial dimensions. Electromagnetic forces are confined on the brane spacetime, while gravitational forces propagate also in the directions external to the brane.}
 \label{fig4}
\end{figure}

In a string theory context, however, the confinement mechanism is not equally efficient for all fundamental interactions. Gravity, in particular, is not confined, or is only partially \cite{27} confined, so that it can propagate  outside the brane spacetime. This possibility is illustrated in Fig. \ref{fig4}, which shows a brane spacetime $\Sg_4$ with two possible sources of interactions. One is a charge, source of the electromagnetic field: the associated electromagnetic waves (or photons) are strcicly confined and can propagate only on $\Sg_4$. The other is a mass, source of the gravitational field: the associated gravitational waves (or gravitons) can leave the brane spacetime $\Sg_4$ and propagate through the external spatial dimensions.

This property of the gravitational field is quite important because, if the higher-dimensional bulk spacetime contains two (or more) fundamental branes, they can interact among themselves gravitationally. And this possibility leads us to an interesting geometric interpretation of the big bang mechanism, namely of the high-energy process which has marked the beginning of the standard cosmological phase, bringing the Universe to the form we are presently observing.

In fact, during the high-curvature phase localized around $t=0$ (Fig. \ref{fig3}), a higher-dimensional Universe tends to be filled by branes which are spontaneously produced in pairs from the high-energy vacuum, and which can gravitationally (and strongly) interact among themselves \cite{19}. According to string theory, on the other hand, the total gravitational force in a higher-dimensional spacetime includes various components: we should mention, in particular, the symmetric-tensor contribution associated to the {\em graviton}, the scalar contribution associated to the {\em dilaton}, and the antisymmetric-tensor contribution associated to the {\em axion} (see, e.g., \cite{18,19}).
\index{axion forces}
\index{brane interactions}

The first two types of forces are always attractive, while the axion force is repulsive between sources of the same sign and attractive between sources of opposite sign (like, for instance, a brane and an antibrane, characterized by opposite axionic charges). It follows, in particular, that if we have two identical branes (or antibranes) in an initial static and symmetric state, then the axion repulsion exactly cancels the attraction due to the graviton and to the dilaton, and the net resulting force is vanishing. If we have instead a brane and an antibrane then the total gravitational force is always non-vanishing and attractive, quite irrespective of their initial configuration.

\begin{figure}[t]
\centering
\includegraphics[height=7cm]{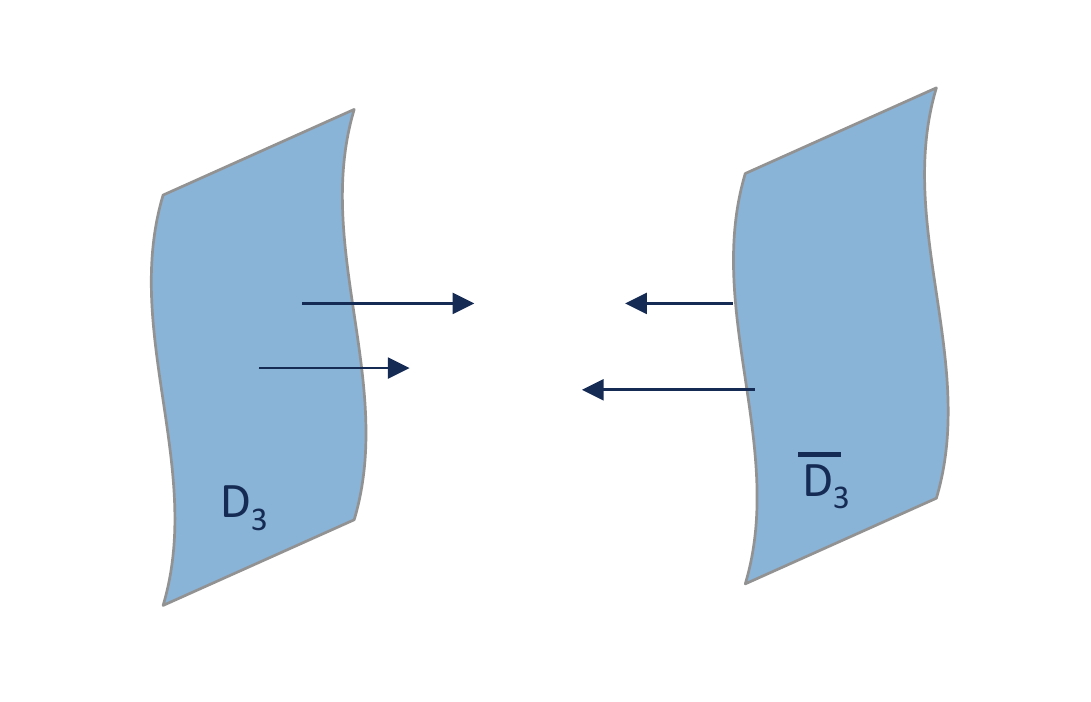}
\caption{\em A brane ($D_3$) and an antibrane ($\overline D_3$) tend to collide because the mutual gravitational force they experience in a higher dimensional spacetime is always of attractive type (like the electric force acting between a charged particle and the corresponding antiparticle in the usual three-dimensional space).}
 \label{fig5}
\end{figure}

\index{anti-brane}
\index{brane collision}
Because of such attractive force, branes and antibranes, copiously produced during the high-energy pre-big bang phase, tend to collide among themselves (Fig. \ref{fig5}): it could be, therefore, that it was the collision of our brane-Universe with an antibrane to simulate the big bang explosion, and trigger the transition from the pre-big bang phase to the phase of standard (post-big bang) evolution. This type of scenario is very similar to the so-called  {\em ekpyrotic} model (first proposed in \cite{28}, and later embedded in the context of a more general type  of {\em cyclic} cosmologies, see e.g. \cite{29}), with the only difference that, in the ekpyrotic case, the $3$-branes are {\em domain walls} representing the spacetime boundaries.
\index{ekpyrotic scenario}

\section{Conclusion}

String theory, M-theory, and the related models of brane interactions  
suggests new and interesting scenarios for the birth of the Universe and its subsequent primordial evolution, not necessarily limited in time by a big bang singularity. They can be tested by present (or near-future) observations concerning the properties -- in particular, the ``blue'' tilt \cite{29a} of the spectrum --  of the cosmic background of relic gravitational radiation (see, e.g., \cite{30}), the possibility of axion contributions to the CMB anisotropies (see, e.g., \cite{30a}), the evolution of the so-called dark energy (or quintessence) field dominating the large-scale dynamics (see, e.g., \cite{31, 31b}).

\index{string-gas cosmology}
\index{brane-gas cosmology}
Some of those scenario have been briefly introduced and illustrated in the previous sections. But there are also other, equally interesting scenarios closely related to the previous ones, among which I would like to mention the string-gas \cite{32} and brane-gas \cite{33} cosmologies, based on the repulsive mechanism of winding modes, as well as more general bouncing cosmology models (see, e.g., \cite{34,34b,35}). Also, models of brane anti-brane inflation \cite{36,37}, where the (time-varying) distance between the two branes plays the role of the inflation field.
\index{bouncing cosmologies}
 
All these models have many (and interesting) phenomenological implications, but -- as usual in a cosmological context -- many studies and many observational data are required before being able of selecting the model most appropriate to our Universe. Thus, we can easily predict that we still have in front of us many years of work and -- maybe -- of surprising findings.

\end{document}